\documentstyle[epsfig,rotating,float,aps,preprint,local]{revtex}
\tightenlines
\begin{document}
\setcounter{figure}{0}

\title{\bf The Cosmic Horizon}
\author{\bf Fulvio Melia\footnote{Sir Thomas Lyle Fellow and Miegunyah Fellow.}}
\address{Department of Physics and Steward Observatory, \\
The University of Arizona,\
Tucson, Arizona 85721, USA\ }
\maketitle
\def\ref{\par\vskip 12pt \noindent \hangafter=1 \hangindent=22.76pt}
\begin{abstract}
The cosmological principle, promoting the view that the universe is homogeneous
and isotropic, is embodied within the mathematical structure of the Robertson-Walker
(RW) metric. The equations derived from an application of this metric to the
Einstein Field Equations describe the expansion of the universe in terms of
comoving coordinates, from which physical distances may be derived using a
time-dependent expansion factor. These coordinates, however, do not explicitly
reveal properties of the cosmic spacetime manifested in Birkhoff's theorem and
its corollary. In this paper, we compare two forms of the metric---written in
(the traditional) comoving coordinates, and a set of observer-dependent
coordinates---first for the well-known de Sitter universe containing only
dark energy, and then for a newly derived form of the RW metric, for a universe
with dark energy and matter. We show that Rindler's event horizon---evident in
the co-moving system---coincides with what one might call the ``curvature
horizon" appearing in the observer-dependent frame. The advantage of this
dual prescription of the cosmic spacetime is that with the latest WMAP results,
we now have a much better determination of the universe's mass-energy content,
which permits us to calculate this curvature with unprecedented accuracy. We use
it here to demonstrate that our observations have probed the limit beyond which
the cosmic curvature prevents any signal from having ever reached us. In the case
of de Sitter, where the mass-energy density is a constant, this limit is fixed for
all time. For a universe with a changing density, this horizon expands until 
de Sitter is reached asymptotically, and then it too ceases to change.
\end{abstract}
\vskip 0.3in

\newpage
\section{\bf Introduction}
The smoothness of the cosmic microwave background (CMB) provides direct evidence that
the universe is isotropic about us. Invoking the Copernican principle---that we do not
live in a special place---allows one to argue further that the universe should be isotropic
around every point, i.e., that it is also (on average) homogeneous. Standard cosmology is therefore
based on the Robertson-Walker (RW) metric for a spatially homogeneous and isotropic
three-dimensional space, expanding or contracting as a function of time:
$$
ds^2=c^2\,dt^2-a^2(t)[dr^2(1-kr^2)^{-1}+
$$
$$
r^2(d\theta^2+\sin^2\theta\,d\phi^2)]\;.
\eqno(1)
$$
\vskip 0.1in\noindent
The coordinates for this metric have been chosen so that $t$ represents the proper time as measured by a
comoving observer, $a(t)$ is the expansion factor, and $r$ is an appropriately scaled radial coordinate
in the comoving frame. The geometric factor $k$ is $+1$ for a closed universe, $0$ for a
flat, open universe, or $-1$ for an open universe.

Assuming that general relativity (GR) provides the correct framework for interpreting cosmological
dynamics, one may then apply the RW metric to Einstein's Field Equations (EFE) to obtain the
(FRW) differential equations of motion. These are the Friedman equation, written as
$$
H^2\equiv\left({\dot a\over a}\right)^2={8\pi G\over 3c^2}\rho-{kc^2\over a^2}\;,
\eqno(2)
$$
and the ``acceleration" equation,
$$
{\ddot a\over a}=-{4\pi G\over 3c^2}(\rho+3p)\;.\eqno(3)
$$
An overdot denotes a derivative with respect to cosmic time $t$, and $\rho$ and $p$
represent the total energy density and total pressure, respectively. A further application
of the RW metric to the energy conservation equation in GR yields the final equation,
$$
\dot\rho=-3H(\rho+p)\eqno(4)
$$
which, however, is not independent of equations (2) and (3).

But written in this form, equations (2), (3), and (4) do not explicitly reveal
the spacetime curvature most elegantly inferred from the \emph{corollary} to
Birkhoff's theorem (Birkhoff 1923). This theorem states that in a spherically
symmetric spacetime, the only solution to the Einstein equations in vacuum is
the Schwarzschild exterior solution. Furthermore, a spherically symmetric
vacuum solution in the exterior is necessarily static. Birkhoff's purpose
was to prove that for general relativity, as with the Newtonian theory, the
exterior gravitational field of a spherically symmetric distribution of matter
is independent of interior radial pulsations.

However, what is relevant to our discussion here is not so much the
Birkhoff theorem per se, but rather its very important corollary. The
latter is a generalization of a well-known result of Newtonian theory
(which also finds a direct application in electrodynamics), in which
the gravitational field of a spherical shell vanishes inside the shell.
The corollary to Birkhoff's theorem states that the metric inside an
empty spherical cavity, at the center of a spherically symmetric system,
must be equivalent to the flat-space Minkowski metric. Space must be
flat in a spherical cavity even if the system is infinite. It matters
not what the constituents of the medium outside the cavity are, as long
as the medium is spherically symmetric.

If one then imagines placing a spherically symmetric mass at the center
of this cavity, according to Birkhoff's theorem and its corollary, the
metric between this mass and the edge of the cavity is necessarily of
the Schwarzschild type. Thus the worldlines linked to an observer in
this region are curved relative to the center of the cavity in a
manner determined solely by the mass we have placed there. This situation
may appear to contradict our assumption of isotropy, which one might
naively take to mean that the spacetime curvature within the medium
should cancel since the observer sees mass-energy equally distributed
in all directions. In fact, the observer's worldlines are curved in
every direction because, according to the corollary to Birkhoff's
theorem, only the mass-energy between any given pair of points in
this medium affects the path linking those points. This is why, of
course, the universe does not expand at a constant rate, due to
the spacetime curvature induced (in every direction) by its internal
constituents.

This consequence of the corollary to Birkhoff's theorem is so
important---and critical to the discussion in this paper---that
it merits re-statement: the spacetime curvature of a wordline
linking any point in the universe to an observer a distance $R$ away
may be determined by calculating the mass-energy enclosed within a
sphere of radius $R$ centered at the origin (i.e., at the location of
the observer). The mass-energy outside of this volume has a net zero
effect on observations made within the sphere.

In this paper, we will introduce a new set of coordinates
(distinct from the co-moving coordinates $[r,t]$), that elicit this
effect directly from the transformed metric, originally appearing
in equation (1). We will show that, with the recent WMAP results
(Spergel et al. 2003), our observational limit clearly corresponds
to the distance beyond which the spacetime curvature prevents any
signal from ever reaching us.

\section{The Asymptotic Universe: de Sitter's Cosmology}
Since the universe appears to be expanding indefinitely, the dark
energy (with constant density) will eventually dominate $\rho$
completely. Other contributions---from radiation and (luminous and
non-luminous) matter---drop off rapidly with increasing volume.

The de Sitter cosmology (de Sitter 1917) corresponds to a universe devoid
of matter and radiation, but filled with a cosmological constant, whose
principal property is the equation of state (EOS) $p=-\rho$. The RW metric in
this case may be written
\vskip -0.05in
$$
ds^2=c^2\,dt^2-e^{2Ht}[dr^2+r^2(d\theta^2+\sin^2\theta\,d\phi^2)]\;,
\eqno(5)
$$
\vskip 0.05in\noindent
where clearly $k=0$ and the expansion factor has the specific form
$a(t)\equiv \exp(Ht)$, in terms of the Hubble constant $H$. (Note that,
though this cosmology may represent the universe's terminal state, it may
also have corresponded to the early inflationary phase, where it would have
produced an exponentiation in size due to the expansion factor $\exp[Ht]$.)

It is not obvious from the form of this metric how the corollary to Birkhoff
theorem's may be recovered. But we must remember that $t$ and $r$ appearing
here are the local (comoving) coordinates. To compare distances and times
over large separations, one must also know $a(t)$. An alternative approach
not used before (or at best only used very rarely) is to introduce the
transformation relating these coordinates to those in the observer's frame.
Only then does the spacetime curvature emerge directly.

For pedagogical purposes, it may be useful here---in understanding the roles
played by our two coordinate systems---to find an analogy between them and those
commonly used in the case of a gravitationally-collapsed object. The comoving
coordinates (including the cosmic time) play the same roles as those of an
observer falling freely under the influence of that object, whereas our new
coordinates correspond to an ``accelerated" frame, like that of an observer
held at a fixed spatial point in the surrounding spacetime. There is an important
difference, however, in that the new set of coordinates $(cT,R,\theta,\phi)$
we introduce below are observer dependent. They are not universal, nor do
they need to be.

Let us now consider de Sitter's metric in its \emph{originally} published form:
$$
ds^2=c^2\,dT^2[1-({R/ R_0})^2]-dR^2[1-({R/ R_0})^2]^{-1}-
$$
$$\qquad\qquad\qquad R^2(d\theta^2+\sin^2\theta\,
d\phi^2)\;.\eqno(6)
$$
The inspiration for this metric is clear upon considering Schwarzschild's (vacuum)
solution describing the spacetime around an enclosed, spherically-symmetric object
of mass $M$:
$$
ds^2=c^2\,dT^2[1-{2GM/ c^2R}]-dR^2[1-
{2GM/ c^2 R}]^{-1}-
$$
$$\qquad\qquad\qquad R^2(d\theta^2+\sin^2\theta\,
d\phi^2)\;.\eqno(7)
$$
\vskip 0.05in\noindent
De Sitter's metric describes the spacetime around
a radially-dependent enclosed mass $M(R)$, as we have for a uniform density
$\rho$ permeating an infinite, homogeneous medium. In that case,
$$
M(R)=M(R_0)({R/ R_0})^3\;,
\eqno(8)
$$
\vskip 0.05in\noindent
and the Schwarzschild factor $[1-({2GM/c^2R})]$ transitions into
$[1-({R/R_0})^2]$. It should be emphasized that equation (6) implicitly
contains the restriction that no mass-energy outside of $R$ should
contribute to the gravitational acceleration inside of $R$, as required by the
corollary to Birkhoff's theorem.

For a given interval $s$, the observer-dependent time $T$ clearly diverges
as $R$ approaches $R_0$. This is the limiting distance beyond which the
spacetime curvature prevents any signal from ever reaching us; the
quantity $R_0$ is the only (classical) scale in the system, and it should not
surprise us---in view of the corollary to Birkhoff's theorem---that it is
defined as a {\it Schwarzschild} radius:
$$
{2GM(R_0)\over c^2}=R_0\;.\eqno(9)
$$
\vskip 0.05in\noindent
That is, $R_0$ is the distance at which the enclosed mass is sufficient to
turn it into the Schwarzschild radius for an observer at the origin of
the coordinates.

It is trivial to show that, in terms of $\rho$, we must have
$$
R_0=\left({3c^4\over 8\pi G\rho}\right)^{1/2}\;,\eqno(10)
$$
\vskip 0.05in\noindent
or more simply, $R_0=c/H$. As it should be, this is also the radius
corresponding to Rindler's event horizon (Rindler 1956), which we will discuss
below. The impact of $R_0$, and hence the corollary to Birkhoff's theorem,
may be gauged directly by considering the transformation of coordinates
$(cT,R,\theta,\phi)\rightarrow (ct,r,\theta,\phi)$, for which
$$
R=e^{Ht}r=e^{ct/R_0}r\;,\eqno(11)
$$
and
$$
T=\ln[\exp(-2Ht)-({r/R_0})^2]^{-1/2H}\;.\eqno(12)
$$
\vskip 0.05in\noindent
It is clear from the definition of $a(t)=\exp(Ht)$, and the metric
coefficients in equation (6), that $R$ is the expanded (or physical) radius,
and $T$ is the time corresponding to $R$ as measured by an observer at the
origin of the coordinates---not the local time measured by a comoving
observer at $r$ (which we have also called the cosmic time $t$). An observer's
worldline must therefore always be restricted to the region $R<R_0$, i.e.,
to radii bounded by the cosmic horizon, consistent with the corollary to
Birkhoff's theorem.

\section{A Universe with Dark Energy and Matter}
But though the contribution from radiation has by now largely subsided,
the universe does contain matter, in addition to dark energy. Unfortunately,
the de Sitter metric is not a good representation of this cosmology. However,
the impact of a cosmic horizon should be independent of the actual metric
used to describe the spacetime. The restrictions on an observer's worldlines
should be set by the physical radius $R_0$, beyond which no signal can reach
her within a finite time, no matter what internal structure the
spacetime may possess.

There now exists empirical evidence that our observation of the early universe
is already close to this limit. The precision measurements (Spergel et al. 2003)
of the CMB radiation indicate that the universe is extremely flat (i.e.,
$k=0$). Thus $\rho$ is at (or very near) the ``critical" value $\rho_0\equiv
3c^2H_0^2/8\pi G$. The Hubble Space Telescope Key Project (Mould et al. 2000)
on the extragalactic distance scale has measured $H$ with unprecedented accuracy,
finding a current value $H_0\equiv H(t_0)=71\pm6$ km s$^{-1}$ Mpc$^{-1}$. (Throughout
this paper, subscript ``0" denotes values pertaining to the current cosmic time $t_0$.)
It is straightforward to show that with this $H_0$, $R_0\approx ct_0$ (specifically,
$13.4$ versus $13.7$ billion lightyears). Note that this $t_0$ is the same as $T$
for an observer at the origin. However, even though $T$ changes
with radius $R$ (see equation 12), the difference between $t$ and $T$ emerges
only when $R\rightarrow R_0$. Thus, as long as $R<R_0$, $t$ and $T$ remain
close to each other. We will consider the properties of $T(R,t)$ for a universe
containing both matter and dark energy shortly.

The near equality $R_0\approx ct_0$ may be an indication that by now our worldlines
are already bounded by the cosmic horizon. It is trivial to see from the FRW
equations that the condition $R_0\approx ct_0$ is equivalent to the statement
that $\ddot a\approx 0$. Type Ia supernova data clearly do not support a
decelerating universe (Perlmutter et al. 1999, Riess et al. 2004), pointing
instead to a universe that is currently coasting ($\ddot a=0$) or, more likely,
one that has undergone periods of past deceleration and present acceleration.
(This should not be confused with the Milne universe (Milne 1940), which itself expands
at a constant rate, though is completely empty with $\rho=0$; such a cosmology is
not relevant to this discussion.) It is easy to write down the appropriate RW
metric when $R_0$ is strictly equal to $ct_0$, for then $\ddot a=0$,
and therefore
\vskip -0.05in
$$
ds^2=c^2dt^2-(H_0t)^2[dr^2+r^2(d\theta^2+\sin^2\theta\, d\phi^2)]\;,\eqno(13)
$$
\vskip 0.05in\noindent
with an expansion factor $a(t)=H_0t$. This is the form of the RW metric we will
use in the rest of this paper, albeit understanding that $R_0$ may not be exactly
equal to $ct_0$ currently. The latter would be realized if $\ddot a/a$ were slightly
greater than zero, as suggested by observations of Type Ia supernovae. But finding
a metric appropriate for a universe containing both matter and dark energy is
very difficult when $\ddot a/a\not=0$, forcing us to adopt this approximation
in order to extend our analysis beyond simply de Sitter. Fortunately, our
results will not depend on this simplification, since we will confirm that a
cosmic horizon emerges in both cases; it is a general property of cosmological
models, independent of the actual spacetime.

The difficult part is to find a transformation, analogous to equations (11) and
(12), that will permit us to write the metric in terms of physical coordinates
($R$ and $T$). The main hurdle here is that, whereas $\rho$ is constant in a de
Sitter universe and no velocity relative to $\rho$ is discernible (so that the
four-velocity in the stress-energy tensor has just a single non-vanishing
component), quite the opposite is true when $\rho$ changes with expansion.

Recognizing that
$$
R=(H_0t)r\;,\eqno(14)
$$
\vskip 0.05in\noindent
we transform the RW metric (equation 13) into the form
$$
ds^2=c^2dt^2[1-(R/ct)^2]+(2R/t)dt\,dR-dR^2
$$
\vskip -0.1in
$$
-R^2(d\theta^2+\sin^2\theta\, d\phi^2)\;.\eqno(15)
$$
\vskip 0.1in\noindent
Already one begins to see the role played by $ct$, which here becomes the
cosmic (or ``curvature") horizon analogous to $R_0$ in equation (6). To diagonalize
the metric and complete the transformation, we put
\vskip -0.05in
$$
dT= \exp[-(1/2)(R/ct)^2]\{dt+dR\,(R/c^2t)
$$
\vskip -0.08in
$$
[1-(R/ct)^2]^{-1}\}[(R/ct)^2-1]^{-1}\;,\eqno(16)
$$
\vskip 0.1in\noindent
which produces the final form,
\vskip -0.05in
$$
ds^2=c^2dT^2[1-(R/ct)^2]^3\exp(R/ct)^2-
$$
\vskip -0.13in
$$
dR^2[1-(R/ct)^2]^{-1}-R^2(d\theta^2+\sin^2\theta\,d\phi^2)\;.\eqno(17)
$$
\vskip 0.08in

Unlike the situation with the de Sitter metric, however, this $dT$ is not an exact
differential. Physically, this simply means that the time difference between two
spacetime points, according to the observer at the origin, depends on the path of
integration. This should not be surprising given that diverse paths sample different
expansion rates as seen in the observer's frame. (Contrast this with the RW metric,
in which the cosmological time $t$ is the same everywhere.)

But since $dT$ depends on only two variables, there exists an integrating factor
(call it $\tau[R,t]$) that converts $dT$ into an exact differential. To find it,
let us define the functions $A(R,t)$ and $B(R,t)$ such that equation (16) may be
written in the form
$$
dT=A(R,t)\,dt+B(R,t)\,dR\;.\eqno(18)
$$
\vskip 0.05in\noindent
In the $R-t$ plane, curves of constant $T$ are specified by the condition
$dT=0$, or equivalently, $dR/dt=-A/B$, which provides the simple solution
\vskip -0.03in
$$
\chi(R,t)\equiv 2\ln (t/t_0)+(R/ct)^2=\chi_0\;.\eqno(19)
$$
\vskip 0.1in\noindent
We have chosen the (integration) constant multiplying $t$ to yield a value
$\chi_0=0$ for $R=0$ at the current time. The constant $\chi_0$ represents the value
of $T$ on the isochronal curve. In terms of $\chi$, the integration factor $\tau(R,t)$
may then be found from the equations
$$
A(R,t)=\tau(R,t){\partial\chi\over\partial t}\;,\eqno(20)
$$
and
$$
B(R,t)=\tau(R,t){\partial\chi\over\partial R}\;.\eqno(21)
$$
Evidently,
$$
\tau(R,t)=-{1\over 2}t\,\left[1-\left({R/ct}\right)^2\right]^{-2}\times\qquad\qquad\qquad
$$
$$
\qquad\qquad\qquad\exp\left[-(1/2)\left({R/ct}\right)^2\right]\;,\eqno(22)
$$
\vskip 0.05in\noindent
and the exact differential representing the time in this coordinate
system is
$$
d\chi={dT\over \tau}\;,\eqno(23)
$$
\vskip 0.05in\noindent
which allows us to migrate from one isochronal curve to the next.

Finally, an alternative form of the metric in equation (17) may be obtained
by replacing $dT$ with the exact differential $d\chi$, using equation (23). The result is
\vskip -0.0in
$$
ds^2=c^2(t\,d\chi)^2\{4[1-(R/ct)^2]\}^{-1}-
$$
\vskip -0.18in
$$
dR^2 [1-(R/ct)^2]^{-1}-R^2(d\theta^2+\sin^2\theta\,d\phi^2)\;.\eqno(24)
$$
\vskip 0.06in

The derivation of the metric in equations (17) and (24) is one of the principal results of
this paper. As was the case with de Sitter, we see directly through its coefficients
the fact that the time $T$ diverges at a specific radius, here equal to $ct=cT(0)$---the
cosmic horizon for this particular spacetime. No wordlines are permitted with $R$
extending beyond this limiting radius. For this reason, it is reasonable to conclude
that the observed near equality $R_0\approx ct$ is not a coincidence, but is instead
an indication that our observational limit has reached the universe's cosmic horizon.

\section{Concluding Remarks}
The cosmic horizon we have discussed in this paper coincides with the event horizon
derived earlier by Rindler (1956) because they represent aspects of the same spacetime,
though viewed in two different coordinates systems. Rindler formalized the
definition of two horizons in cosmology, both functioning as hypersufaces in spacetime
that divide ``things" into two separate non-null classes. For a fundamental observer
$A$, the {\it event} horizon is defined to be a hypersurface that divides all events
into the class in which these have been, are, or will be, observable by $A$, and the
complementary class in which the events are forever outside of $A$'s power of observation.
These event horizons are created by the universe's expansion, which produces distant
regions receding from $A$ at such speeds that no causal contact can ever occur between
them and the observer. Since the comoving coordinates represent the universe's
``freely falling" frame, it is clear that this critical condition must therefore
correspond to the radius $R_0$ at which the curvature in the cosmic spacetime causes
$T$ to diverge.

Rindler's second type of horizon is called a {\it particle} horizon. For the
fundamental observer $A$ and cosmic instant $t_0$, the particle horizon is a
surface in instantaneous 3-space at time $t=t_0$ that divides all fundamental
particles into two classes: those that have already been observed by $A$ at
time $t_0$, and those that have not. These horizons are therefore created
by the finite propagation speed of signals that couple particles at distance
$d$ to the observer $A$ for times $t>d/c$. The cosmic horizon $R_0$ is not
related to the particle horizon.

An interesting property of $R_0$ emerges upon closer scrutiny of its value in
the case of de Sitter and the alternate cosmology containing both matter and
dark energy. First of all, the density $\rho$ is constant in the former, so
$R_0$ does not change with time. Thus, the horizon in de Sitter is fixed forever,
and no events occurring beyond it can ever be in causal contact with the observer
at the origin. This is presumably the situation that emerges as the universe
approaches de Sitter asymptotically, in which limit $R_0$ must be 
calculated for $\rho$ due to dark energy alone. Second, the density $\rho$
in the alternate metric decreases with time and therefore $R_0$ correspondingly 
increases---i.e., $R_0(t)$ is a function of $t=T(0)$. This means, of course,
that some portions of the universe that produced effects at $R<R_0(t)$ we 
measure at time $t$ have by now moved beyond the limiting radius. However,
the effects of gravity travel at the speed of light, so what matters in 
setting the structure of the universe within the horizon at time $t$ is the
mass-energy content within $R_0$. The influence of these distant regions
of the universe ended once their radius from us exceeded $R_0$.

The light-travel time distance ($\sim$13.7 billion light-years) could only have been
identified with the cosmic horizon once the data (Spergel et al. 2003) confirmed
that the universe is flat, and revealed that the density $\rho$ is close to, or at,
its ``critical" value. It has never been quite clear why this critical condition
exists, except for the possibility that inflation in the early universe could have
led to this. But based on our analysis of the various metrics in this paper, an
alternative interpretation---or perhaps simply an additional reason---for this criticality
is that the universe has now expanded long enough for the condition $R_0\approx ct$
to have been attained.

One can easily show from the FRW equations that once $R_0$ has been reached,
the expansion thereafter proceeds at a constant rate. But this does not necessarily
mean that this dynamical state must have been present since the Big Bang, so it is
by no means trivial to see how such a scenario would affect Big Bang nucleosynthesis
(Burles et al. 2001), and structure formation in the early universe (Springel et al.
2006).

But for the current universe, the near-equality $R_0\approx ct$ would produce an observable
signature because it is equivalent to the condition $\ddot a\ge 0$. At least out to
a redshift of $\sim 1.8$, the universe appears to be expanding with a slight positive
acceleration (Riess et al. 2004). With ongoing observations of Type Ia supernovae, the
value of $\ddot a$ will continue to be refined and, with it, so too the value of $R_0-ct$.

Clearly, more work needs to be done. There is little doubt that a cosmic horizon
exists. It is required by the application of the corollary to Birkhoff's theorem
to an infinite, homogeneous medium, and there is some evidence that we have already
observed phenomena close to it. However, it may be that observational cosmology
is not entirely consistent with the condition $R_0\approx ct$ in the current epoc.
If not, there must be some other reason for this apparent coincidence. Perhaps the
assumption of an infinite, homogeneous universe is incorrect. Whatever the case may
be, the answer could be even more interesting than the one we have explored here.

\section*{Acknowledgments}
This research was partially supported by NSF grant 0402502 at the
University of Arizona, and a Miegunyah Fellowship at the University
of Melbourne. Helpful discussions with Ray Volkas and Hideki Maeda
are greatly appreciated. I am particularly grateful to Roy Kerr for
many inspirational discussions. In addition, the anonymous referee
provided a thoughtful analysis that has led to an improvement in
presentation. Part of this work was carried out at the Center for 
Particle Astrophysics and Cosmology in Paris.

\end{document}